\documentclass[twocolumn,openany,openright,amsmath,amssymb,superscriptaddress]{revtex4-1}

\usepackage{float}
\usepackage{latexsym} 
\usepackage{amsmath,amsthm}
\usepackage{ifpdf}
\usepackage{epstopdf}
\usepackage{subfig}
\usepackage{soul}
\usepackage{dcolumn}
\usepackage{bm}
\usepackage{braket}
\usepackage{wrapfig}
\usepackage[dvipsnames]{xcolor}
\usepackage{color}
\usepackage{hyperref}

\usepackage{graphicx}

\begin{document}
\newcommand{\umlaut}[1]{\ddot{\textrm{#1}}}
\newcommand{\beq}{\begin{equation}}
\newcommand{\eeq}{\end{equation}}
\newcommand{\barr}{\begin{eqnarray}}
\newcommand{\earr}{\end{eqnarray}}
\newcommand{\bseq}{\begin{subequations}}
\newcommand{\eseq}{\end{subequations}}
\newcommand{\oper}[1]{\hat{#1}}
\newcommand{\adj}[1]{\oper{#1}^{\dagger}}
\newcommand{\ann}{\hat{a}}
\newcommand{\creat}{\hat{a}^{\dagger}}
\newcommand{\commutator}[2]{[\oper{#1},\oper{#2}]}
\newcommand{\outerP}[2]{|#1\rangle\langle#2|}
\newcommand{\expectation}[3]{\langle #1|#2|#3\rangle}
\newcommand{\proiettore}[1]{\ket{#1}\bra{#1}}
\newcommand{\closure}[1]{\sum_{#1}\proiettore{#1}=1}
\newcommand{\closureInt}[1]{\int d#1\proiettore{#1}=1}
\newcommand{\vett}[1]{\textbf{#1}}
\newcommand{\uvett}[1]{\hat{\textbf{#1}}}
\newcommand{\derivata}[2]{\frac{\partial #1}{\partial #2}}

\newcommand{\creak}[1]{\hat{a}^{\dagger}_{#1}}
\newcommand{\annak}[1]{\hat{a}_{#1}}
\newcommand{\crebk}[1]{\hat{b}^{\dagger}_{#1}}
\newcommand{\annbk}[1]{\hat{b}_{#1}}

\newcommand{\crek}[2]{\hat{#1}^{\dagger}_{#2}}
\newcommand{\annk}[2]{\hat{#1}_{#2}}
\newcommand{\red}[1]{{\color{red}{#1}}}
\newcommand{\blue}[1]{{\color{blue}{#1}}}

\title{Inhomogeneous Dirac-Bloch Equations for Graphene Interacting with Structured Light}
\author{Yaraslau Tamashevich}
\affiliation{Faculty of Engineering and Natural Sciences, Tampere University, Tampere 33720, Finland}
\author{Marco Ornigotti}
\affiliation{Faculty of Engineering and Natural Sciences, Tampere University, Tampere 33720, Finland}

\begin{abstract}
We generalise the usual framework of Dirac-Bloch equations, used to compute the nonlinear optical response of 2D materials excited by spatially uniform optical pulses, to the case of structured light pulses. We derive the general form of Dirac-Bloch equations in the presence of a spatially inhomogeneous field, for the case of a graphene-like material. Then, as as example of application of our method, we consider explicitly the case of angular momentum carrying optical pulses interacting with graphene. 
\end{abstract}

\maketitle

\section{\label{sec:Intoduction}Introduction}
Since the discovery of graphene \cite{novoselov_electric_2004} a lot of effort in theoretical and experimental works has been done in order to describe different properties of graphene. 
This material exhibit exotic electronic and optical properties such as minimal conductivity \cite{refCond1}, universal absorbance \cite{univAbs1}, anomalous Quantum Hall Effect \cite{novoselov_qhe,zhang_qhe}, and high nonlinear optical response \cite{wright_strong_2009,hafez_terahertz_2020}, to name a few.
Inspired by the new physics discovered in graphene, an intensive research has been carried out in the last decade on other 2D materials as well, such as transition metal dichalcogenides (TMDs) \cite{manzeli_2d_2017,autere_nonlinear_2018},hexagonal boron nitride (hBN) \cite{neto2009electronic,caldwell_photonics_2019},  and black phosphorous \cite{blackPh}.
The most interesting characteristics of 2D materials, however, is that they possess high nonlinear response, which can be several orders of magnitude higher than their bulk counterparts. This enhancement allows to observe a lot of nonlinear effects such as, for example, second- \cite{zeng2013optical} and third-harmonic generation \cite{kumar2013third}, four-wave mixing \cite{hendry_coherent_2010}, and optical limiting \cite{dong_optical_2015}. While graphene is intrinsically centrosymmetric and therefore lacks second order response, recent works have demonstrated how it is possible to induce second-harmonic generation in graphene by inducing stress or strains \cite{ornigotti_strain_2021} or by exploiting the fact that an ultra-intense pulse can open its own gap near Dirac points, thus breaking centrosymmetry \cite{fabio1}. 

On a seemingly parallel trail, structured light and, in particular, orbital angular momentum (OAM) carrying beams have been widely studied in the past decades since the pioneering work of Allen and Woerdman in 1992 \cite{allen_orbital_1992}, and represent a versatile tool used in several different experimental situations, and show wide range of application in different areas, namely, optical tweezers \cite{oam1}, quantum information \cite{oam2}, microscopy \cite{oam3}, spectroscopy \cite{oam4}, classical \cite{oam5} and quantum \cite{oam6} communication, to name a few.

Despite the considerable amount of research that has been done in both 2D materials and structured light, investigating the benefits and potential new physics that would result by combining these two disciplines is still in its infancy, and it is only recently, that different scenarios, such as the interaction of OAM beams with ring semiconductors \cite{watzel_centrifugal_2016}, photoexcitation of graphene \cite{farias_photoexcitation_2013} and structured nonlinear optics in 2D materials have been studied. A comprehensive toolbox for analysing the effects of structured light on electron dynamics in 2D materials, however, is still not available. 

In this work, we then extend the usual Dirac-Bloch equation approach \cite{ishikawa} to the case of a spatially inhomogeneous, impinging electromagnetic field and develop a framework that will allow us to calculate nonlinear current and nonlinear optical response of graphene-like materials when excited by structured light.

This work is organized as follows: in Sect. \ref{sec:graphene} we briefly summarize the main results of graphene, in terms of low-energy Hamiltonian, its eigenstates and eigenvalues, which will constitute the building blocks of our approach. Then, in Sect. \ref{sec:intH} we discuss the explicit form of the interaction Hamiltonian, in the minimal coupling approximation, for the case of a structured impinging light pulse, and give the explicit expression for the electric dipole moment of graphene in $k$-space, and express the interaction Hamiltonian in the electron-hole basis. Section \ref{sec:diracBloch} is then dedicated to constructing Dirac-Bloch equations for the case of impinging structured light pulses, while Sect. \ref{sec:current} deals with calculating the Dirac current for this general case. The nonlinear response of graphene under structured light excitation is then discussed in Sect. \ref{sec:nonOptRes}. Finally, conclusions and future perspectives of this work are discussed in Sect. \ref{sec:conc}. 
\section{Low-Energy Hamiltonian for Graphene}\label{sec:graphene}
\begin{figure}[!t]
\begin{center}
\includegraphics[width=0.4\textwidth]{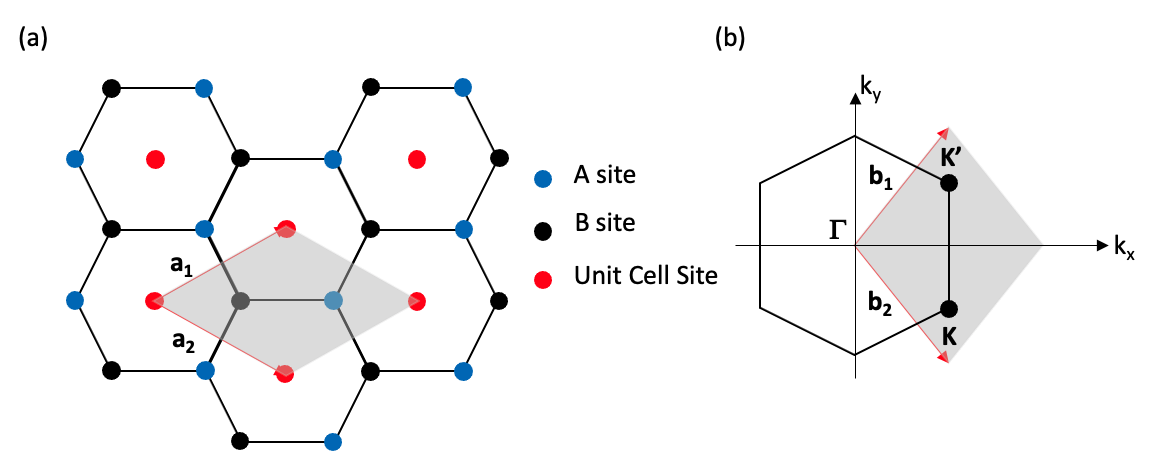}
\caption{(a) Lattice structure of graphene. The unit cell, containing two atoms (one belonging to lattice A, the other to lattice B), is shaded in gray. The basis vectors used in this work are chosen to be $\vett{a}_1=(a/2)(3,\sqrt{3})$, and $\vett{a}_2=(a/2)(3,-\sqrt{3})$. (b) Reciprocal space lattice structure of graphene (the Wigner-Seitz cell is shaded in grey). The reciprocal lattice basis vectors are defined as $\vett{b}_1=(2\pi/3a)(1,\sqrt{3})$ and $\vett{b}_2=(2\pi/3a)(1,-\sqrt{3})$.The two inequivalent Dirac points $\vett{K}$ and $\vett{K'}$ are defined as $\vett{K}=(2\pi/3a)(1,-1/\sqrt{3})$, and $\vett{K'}=(2\pi/3a)(1,1/\sqrt{3})$,respectively.}
\label{figure1}
\end{center}
\end{figure}

%
To derive the low energy Hamiltonian for graphene in the presence of an external field, we can start from the minimally coupled crystalline Hamiltonian
\beq
\hat{H}=\int d^3r\hat{\psi}^{\dagger}(\vett{r})\left[\frac{\left(\hat{\vett{p}}-e\vett{A}\right)^{2}}{2m}+\phi_c(\vett{r})+V_B(\vett{r})\right]\hat{\psi}(\vett{r}),
\eeq
where $\phi_c(\vett{r})$ is the Coulomb potential, $V_B(\vett{r})$ is the crystalline (periodic) potential, $\hat{\psi}(\vett{r})$ is the electron wavefunction operator, and $\vett{A}(\vett{r},t)$ is the vector potential. From here henceforth, we are assuming to work in the Coulomb gauge, where $\nabla\cdot\vett{A}=0$, and we also neglect the Coulomb interaction terms, since they don't contribute significantly for the case of graphene \cite{grapheneBook}. Since for our purposes it is more convenient to work with the electric field, rather than the vector potential, we can apply the Power-Zienau-Wooley transformation \cite{methodsQO} to the above Hamiltonian, to get the usual light-matter interction in the electric dipole approximation as follows:
\beq
\hat{H}=\hat{H}_0+\hat{H}_I,
\eeq
where
\bseq\label{eq4}
\begin{align}
\hat{H}_0&=\int\,d^3r\,\hat{\psi}^{\dagger}(\vett{r})\left[\frac{\hat{\vett{p}}^2}{2m}+V_B(\vett{r})\right]\hat{\psi}(\vett{r}) \\  \nonumber &\equiv\int\,d^3r\,\hat{\psi}^{\dagger}(\vett{r})\mathcal{H}_0\hat{\psi}(\vett{r}),\\
\hat{H}_I&=-e\int\,d^3r\,\hat{\psi}^{\dagger}(\vett{r})\,\left[\vett{r}\cdot\vett{E}_T(\vett{r},t)\right]\,\hat{\psi}(\vett{r}),
\end{align}
\eseq
where $\vett{E}_T(\vett{r},t)=-\partial_t\vett{A}(\vett{r},t)$ is the (transverse) electric field. 
To calculate the graphene Hamiltonian, we use the following Bloch representation for the electron wavefunction operator
\barr
\label{eq6}
\hat{\psi}(\vett{r})=\frac{1}{\sqrt{N}}\sum_{k}\sum_{n=0}^Ne^{i\vett{k}\cdot\vett{R}_n}\Big[\phi_A(\vett{r}-\vett{R}_n)\annak{k} \\ \nonumber
+\phi_B(\vett{r}-\vett{R}_n)\annbk{k}\Big],
\earr
where $\phi_{A,B}(\vett{r})$ are $2p_z$-orbital wavefunctions at site $A$ and $B$ (and they form an orthogonal normalised basis set), $\vett{R}_n=j\vett{a}_1+m\vett{a}_2$ is a Bravais lattice vector (see Fig. \ref{figure1}), and $\annak{k},\annbk{k}$ are the electron site operators, annihilating an electron on site A or B, respectively.
\begin{figure}
\begin{center}
\includegraphics[width=0.45\textwidth]{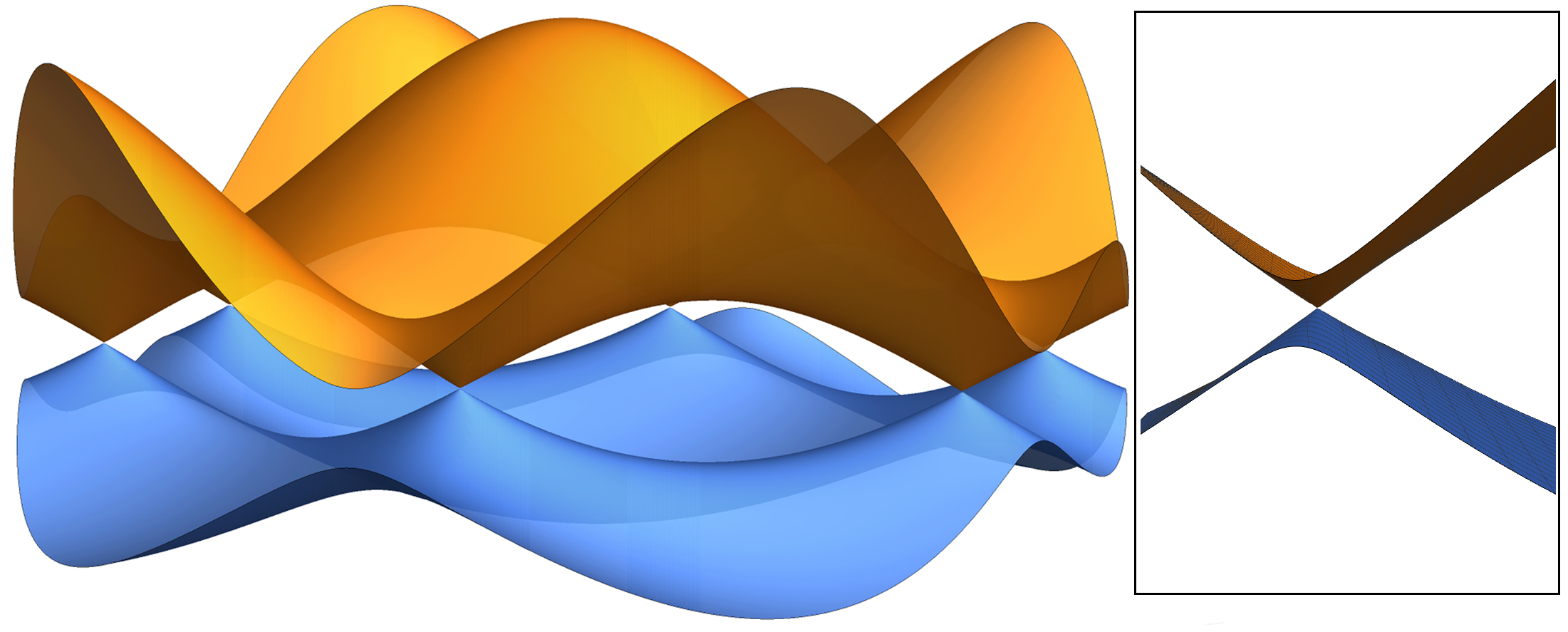}
\caption{Band structure of graphene. You can see 6 touching point of valence (orange) and conduction (blue) band which are called Dirac points. Locally Dirac point can be represented as a touching of two identical cones (see figure inside rectangular) that allows there treating as a massless particles.}
\label{graphene_band}
\end{center}
\end{figure}

By substituting Eq. \eqref{eq6} into the first of Eqs. \eqref{eq4}, it is not difficult to see, that the free Hamiltonian $\hat{H}_0$ can be then reduced to 
\beq
\hat{H}_0=\gamma\sum_k\left[f(\vett{k})\creak{k}\annbk{k}+\text{h.c.}\right],
\eeq
where $\gamma$ is the hopping amplitude and $f(\vett{k})=1+\exp{(i\vett{k}\cdot\vett{a}_1)}+\exp{(i\vett{k}\cdot\vett{a}_2)}$, and $\text{h.c.}$ stands for Hermitian conjugate. Diagonalising the above free Hamiltonian leads to the band structure (eigenvalues) of graphene, i.e., $E(\vett{k})=\pm\gamma|f(\vett{k})|$, and its corresponding eigenvectors
\beq\label{eq18}
u_{\pm}(\vett{k})=\frac{1}{\sqrt{2}}\left(\begin{array}{c}
e^{i\frac{\varphi(\vett{k})}{2}}\\
\pm\,e^{-i\frac{\varphi(\vett{k}}{2})}
\end{array}\right),
\eeq
where $\varphi(\vett{k})=\text{arg}\left[f(\vett{k})\right]$. A plot of the full band structure of graphene is shown in Fig. \ref{graphene_band} 
We can transform the site operators as well into the diagonal basis of the Hamiltonian $\hat{H}_0$ ,obtaining
\beq\label{eq20}
\left(\begin{array}{c}
\annk{e}{k}\\
\crek{h}{-k}
\end{array}\right)=\frac{1}{\sqrt{2}}\left(\begin{array}{cc}
e^{-i\frac{\varphi(\vett{k})}{2}} & e^{i\frac{\varphi(\vett{k})}{2}}\\
e^{-i\frac{\varphi(\vett{k})}{2}} & -e^{i\frac{\varphi(\vett{k})}{2}}
\end{array}\right)\left(\begin{array}{c}
\annk{a}{k}\\
\annk{b}{k}
\end{array}\right).
\eeq
Notice that this is equivalent to the electron-hole representation discussed, for example, in Ref. \cite{koch}, where $\annk{e}{k}$ annihilates an electron with momentum $k$, and $\crek{h}{-k}$ creates a hole with momentum $-k$.

The free Hamiltonian in the electron-hole representation can be then written as follows
\beq
\hat{H}_0=\sum_k\,E(\vett{k})\left(\crek{e}{k}\annk{e}{k}+\crek{h}{-k}\annk{h}{-k}\right).
\eeq

\subsection{Low-Energy Approximation}
We can expand the eigenvalues $E(\vett{k})$ around the zeros of $f(\vett{k})$, which happen to be the two inequivalent Dirac points $\vett{K}$ and $\vett{K'}$ [see Fig. \ref{figure1} (b)] \cite{katsnelson} to obtain the low-energy approximation of $\hat{H}_0$ around the Dirac points. To do that, we first redefine the coordinates of the Dirac points as $\vett{K}=(2\pi/3a)(1,\xi/\sqrt{3})$, where $\xi=\pm 1$, and then expand $f(\vett{k})$ around $\vett{q}=\vett{k}+\vett{K}$ obtaining
\barr
f(\vett{q})&\simeq& f(\vett{K})+\nabla_kf(\vett{k})\Big|_{\vett{K}}\vett{q}\nonumber\\
&=&\frac{3ai}{2}\left(k_x-i\xi\,k_y\right)\nonumber\\
\earr
Substituting this result in the expression of the Hamiltonian $\hat{H}_0$ and defining the Fermi velocity as $\hbar v_f3a\gamma/2$, we get the familiar form of the low-energy graphene Hamiltonian, i.e.,
\beq
\hat{H}(\vett{k})=\hbar v_f(\sigma_xk_x+\xi\sigma_y k_y),
\eeq
where $\xi=\pm1$ is the valley index, and $\sigma_i$ are Pauli matrices. The eigenvalues of $\hat{H}(\vett{k})$ are now simply $E(\vett{k})=\hbar v_f|\vett{k}|$. Notice, moreover, that in this approximation
\barr
\varphi(\vett{k})&=&\text{arg}\left[f(\vett{k})\right]\simeq-\frac{\pi}{2}+\xi\arctan\left(\frac{k_y}{k_x}\right) \\ \nonumber
&\equiv&-\frac{\pi}{2}+\varphi_{\xi}(\vett{k}).
\earr
And, that the eigenvectors $u_{\pm}(\vett{k})$ acquire a valley dependence through $\varphi_{\xi}(\vett{k})$, i.e., $u_{\pm}(\vett{k})\rightarrow u_{\pm}^{\xi}(\vett{k})$.
\section{Interaction Hamiltonian}\label{sec:intH}
We now substitute the Bloch Ansatz given by Eq. \eqref{eq6} into the expression of the interaction Hamiltonian, i.e., the second of Eqs. \eqref{eq4}. To calculate the integrals, we proceed as before, first by introducing the change of variables $\vett{x}=\vett{r}-\vett{R}_n$, assuming only nearest neighbour coupling, such that $\vett{R}_{\ell}=\vett{R}_m-\vett{R}_n$ only allows for $\ell=\{0,\pm1\}$, and we make the extra assumption that the electric field is constant over a single unit cell, i.e.,
\beq
\vett{E}_T(\vett{x}+\vett{R}_n,t)\simeq\,\vett{E}_T(\vett{R}_n,t).
\eeq
Notice that this approximation is consistent with previous work dealing with semiconductor Bloch equations in the presence of inhomogeneous fields \cite{refN}

Since we are doing these calculation in the site basis, we will have to deal with both diagonal terms, proportional to $\creak{k}\annak{k}$ and $\crebk{k}\annbk{k}$, and nondiagonal terms, proportional to $\creak{k}\annbk{k}$. The former are related to the presence of permanent electric dipoles in the system, and without loss of generality we can neglect their contribution, since graphene (and most 2D materials, in general) do not possess permanent dipoles. Therefore, we focus our attention on the nondiagonal interaction terms, which describe light-induced transitions and are regulated by the usual electric dipole moment.
The general form of the (nondiagonal) interaction Hamiltonian is then given by
%
\barr
\hat{H}_I(\vett{k})&=&\sum_ne^{i(\vett{k}-\vett{k'})\cdot\vett{R}_n}\,\Bigg[\vett{E}_T(\vett{R}_n,t)\cdot\vett{d}_{CV}(\vett{k'}) \\ \nonumber
-&e&\,f(\vett{k'})\vett{R}_n\cdot\vett{E}_T(\vett{R}_n,t)\,\int\,\frac{d^3x}{N}\phi_A^*(\vett{x})\phi_B(\vett{x})\Bigg]\crek{a}{k}\annk{b}{k'}\nonumber\\
\earr
where the integral in the second line is zero because $\phi_A(x)$ and $\phi_B(x)$ are orthogonal,
\beq
\vett{d}_{CV}(\vett{k})=-e\sum_{\ell}e^{i\vett{k}\cdot\vett{R}_{\ell}}\int\,d^3x\phi_A^*(\vett{x})\,\vett{x}\,\phi_B(\vett{x}-\vett{R}_{\ell})\nonumber\\
\eeq
is the (conduction-to-valence-band) transition dipole moment in reciprocal space. Notice, that while normally, with spatially homogeneous fields, the first term in the equation above would be readily calculated because $\vett{E}_T(\vett{R}_n,t)=\vett{E}_T(t)$ constant in space, this is not the case here, as we are considering a general, spatially inhomogeneous field. In this case, we can introduce the angular spectrum representation for the impinging field (i.e., its Fourier transform) as 
\beq
\vett{E}_T(\vett{R}_n,t)=\sum_q\mathcal{E}_T(\vett{q},t)e^{-i\vett{q}\cdot\vett{R}_n},
\eeq
and then introduce the generalised Rabi frequency as follows
\beq\label{eq35}
\Omega(\vett{k},\vett{q},t)=\frac{1}{\hbar}\mathcal{E}_T(\vett{q},t)\cdot\vett{d}_{CV}(\vett{k}-\vett{q}),
\eeq
so that the interaction Hamiltonian above can be written as
\beq
\hat{H}_I(\vett{k})=\hbar\sum_q\,\Omega(\vett{k},\vett{q},t)\crek{a}{k}\annk{b}{k-q}.
\eeq
Notice, that in order for $\Omega(\vett{k},\vett{q},t)$ to be a bona-fide Rabi frequency, it must be a dimensionless quantity, which implicitly means that the dimension of the electric field must be $Vs/m$, which is consistent with the fact that $|\vett{E}|=|\partial_t \vett{A}|\simeq\tau A_0$, with $\tau$ being the characteristic timescale of the impinging electric field (for an impinging pulse, $\tau$ is the duration of the pulse).

We are now in the position to write down the total interaction Hamiltonian $\hat{H}_I$ in the site-basis as follows
\beq\label{eq37}
\hat{H}_I=\hbar\sum_{k,q}\left[\Omega(\vett{k},\vett{q},t)\crek{a}{k}\annk{b}{k-q}+\Omega^*(\vett{k},\vett{q},t)\crek{b}{k}\annk{a}{k-q}\right].
\eeq
This is the first result of our work. Notice, how accounting for the spatial variation of the impinging electromagnetic pulse means that we are forced to introduce a nonlocal coupling between the various states in $k$ space, through the operators $\creak{k}\annbk{k-q}$, with the (transverse) momentum of the light pulse acting as the connection between the various $k$ states available in the first Brillouin zone.

The interaction Hamiltonian in Eq. \eqref{eq37} can be put in a simpler and more straightforward form, by representing the (complex) Rabi frequency in polar coordinates, such that we get
\barr\label{eq37bis}
\hat{H}_I=\hbar\sum_{k,q}\left|\Omega(\vett{k},\vett{q},t)\right|\Big[e^{i\omega(\vett{k},\vett{q},t)}\crek{a}{k}\annk{b}{k-q} \\ \nonumber
+e^{-i\omega(\vett{k},\vett{q},t)}\crek{b}{k}\annk{a}{k-q}\Big],
\earr
where $\left|\Omega(\vett{k},\vett{q},t)\right|=\sqrt{\Omega_x^2(\vett{k},\vett{q},t)+\Omega_y^2(\vett{k},\vett{q},t)}$ is the modulus of the Rabi frequency, and 
\beq\label{Rabi_frequency_phase}
\omega(\vett{k},\vett{q},t)=\text{arg}\left[\Omega(\vett{k},\vett{q},t)\right]=\arctan\left[\frac{\Omega_y(\vett{k},\vett{q},t)}{\Omega_x(\vett{k},\vett{q},t)}\right],
\eeq
is its phase.
\subsection{Conduction-to-Valence band Dipole Moment in Reciprocal Space}
The explicit expression of the conduction-to-valence band dipole moment can be calculated analytically, in the electron-hole representation, using the eigenstates (in reciprocal space) of the free Hamiltonian $\hat{H}(\vett{k})$, as detailed in Ref. \cite{ref4} as
\beq
\vett{d}_{CV}(\vett{k})=\expectation{\psi_{c}(\vett{k})}{e\vett{r}}{\psi_{v}(\vett{k})},
\eeq
where the conduction and valence band eigenstates are given as 
$
\ket{\psi_{c,v}(\vett{k})}=e^{i\vett{k}\cdot\vett{r}}\ket{u_{\pm}(\vett{k})},
$
where the $+$ ($-$) sign refers to conduction (valence) band. Substitution gives
\beq
\vett{d}_{CV}(\vett{k})=\frac{e}{2}\nabla_k\varphi(\vett{k}).
\eeq
Notice that in the vicinity of a Dirac point $\varphi(\vett{k})\simeq\frac{\pi}{2}+\xi\arctan(k_y/k_x)$, and the dipole moment becomes
\barr
\vett{d}_{CV}(\vett{k})&=&\frac{e\xi}{2}\left(-\frac{k_y}{k_x^2+k_y^2}\uvett{x}+\frac{k_x}{k_x^2+k_y^2}\uvett{y}\right) \\ \nonumber &=&-\frac{e\xi}{2|\vett{k}|}\left[\sin\varphi_{\xi}(\vett{k})\uvett{x}-\cos\varphi_{\xi}(\vett{k})\uvett{y}\right],
\earr
or, by introducing the helicity basis $\uvett{h}_{\xi}=\frac{1}{\sqrt{2}}\left(\uvett{x}+i\xi\uvett{y}\right)$ \cite{mandelWolf}, we get
\beq
\vett{d}_{CV}(\vett{K})=-\frac{ie\xi}{2\sqrt{2}|\vett{k}|}\left(e^{-i\varphi_{\xi}(\vett{k})}\uvett{h}_+-e^{i\varphi_{\xi}(\vett{k})}\,\uvett{h}_-\right)
\eeq

%
\subsection{Interaction Hamiltonian in the Electron-Hole Representation}
We can transform the interaction Hamiltonian in the electron-hole representation using the inverse of the change of basis given by Eq. \eqref{eq20}, i.e.,
\beq\label{eq38}
\left(\begin{array}{c}
\annk{a}{k}\\
\annk{b}{k}
\end{array}\right)=\frac{1}{\sqrt{2}}\left(\begin{array}{cc}
e^{i\frac{\varphi(\vett{k})}{2}} & e^{i\frac{\varphi(\vett{k})}{2}}\\
e^{-i\frac{\varphi(\vett{k})}{2}} & -e^{-i\frac{\varphi(\vett{k})}{2}}
\end{array}\right)\left(\begin{array}{c}
\annk{e}{k}\\
\crek{h}{-k}
\end{array}\right).
\eeq
Substituting this into Eq. \eqref{eq37}, and using the anti-commutation relations $\{\crek{h}{k},\annk{h}{q}\}=\delta_{k,q}$ we get 
\barr\label{eq44}
\hat{H}_I=\hbar\sum_{k,q}\left|\Omega(\vett{k},\vett{q},t)\right| &\Big\{\cos\Phi(\vett{k},\vett{q},t)\Big(\crek{e}{k}\annk{e}{k-q}+\crek{h}{q-k}\annk{h}{-k}\nonumber\\
-\delta_{q,0}\Big) +i\sin\Phi(\vett{k},\vett{q},t)&\left(\crek{e}{k}\crek{h}{q-k}-\annk{h}{-k}\annk{e}{k-q}\right)\Big\},
\earr
where
\beq\label{eq42}
\Phi(\vett{k},\vett{q},t)=\frac{\Delta\varphi(\vett{k},\vett{q})}{2}-\omega(\vett{k},\vett{q},t),
\eeq
with $\Delta\varphi(\vett{k},\vett{q})=\varphi(\vett{k})+\varphi(\vett{k}-\vett{q})]$, and $\omega(\vett{k},\vett{q},t)$ defined in Eq.\eqref{Rabi_frequency_phase}.
The form of the interaction Hamiltonian in the electron-hole representation is quite complicated, because it contains not only interband transitions ($e^{\dagger} e $ and $ h^{\dagger} h$ terms) but also interband ones ($e^{\dagger} h^{\dagger}$ and $ h e $ terms). Moreover, written in this form it is equivalent to the linear minimal coupling Hamiltonian in Ref. \cite{koch2}.
\section{Dirac-Bloch Equations}\label{sec:diracBloch}
To calculate the Dirac-Bloch equations, we first introduce the following two-body populations and polarisation
\bseq
\begin{align}
n_e(\vett{k},\vett{q})&=\langle\crek{e}{k}\annk{e}{q}\rangle,\\
n_h(\vett{k},\vett{q})&=\langle\crek{h}{-k}\annk{h}{-q}\rangle,\\
p(\vett{k},\vett{q})&=\langle \annk{h}{-k}\annk{e}{q}\rangle,
\end{align}
\eseq
and calculate their time derivative with the help of the Heisenberg equation. Using the anti-commutation rules $\{\crek{e}{k},\annk{e}{k'}\}=\delta_{k,k'}$ and $\{\crek{h}{-k},\annk{h}{-k'}\}=\delta_{k,k'}$ we get, after some long but straightforward algebra, the following two-body Dirac-Bloch equations
\begin{widetext} 
\bseq\label{DiracBloch}
\begin{align}
-i\frac{d}{dt}n_e(\vett{k},\vett{k}')&=\frac{E(\vett{k})-E(\vett{k}')}{\hbar}n_e(\vett{k},\vett{k}') +\sum_q\Bigg\{G_{c}(k+q,q,t)n_e(\vett{k}+\vett{q},\vett{k}')-G_c(\vett{k}',\vett{q},t)n_e(\vett{k},\vett{k}'-\vett{q})\nonumber\\
&-i\left[G_s(\vett{k}',\vett{q},t)p^*(\vett{k}'-\vett{q},\vett{k})-G_s(\vett{k},\vett{q},t)p(\vett{k}',\vett{k}-\vett{q})\right]\Bigg\},\\
-i\frac{d}{dt}n_h(\vett{k},\vett{k}')&=\frac{E(\vett{k})-E(\vett{k}')}{\hbar}n_h(\vett{k},\vett{k}')+\sum_q\Bigg\{G_c(\vett{k},\vett{q},t)n_h(\vett{k}-\vett{q},\vett{k}')-G_c(\vett{k}'+\vett{q},\vett{q},t)n_h(\vett{k},\vett{k}'+\vett{q})\nonumber\\
&+\left[G_s^*(\vett{k}'+\vett{q},\vett{q},t)p^*(\vett{k},\vett{k}'+\vett{q})+G_s(\vett{k},\vett{q},t)p(\vett{k}-\vett{q},\vett{k}')\right]\Bigg\},\\
-i\frac{d}{dt}p(\vett{k},\vett{k}')&=-\frac{E(\vett{k}')+E(\vett{k})}{\hbar}p(\vett{k},\vett{k}')+\sum_q\Bigg\{-G_c(\vett{k}',\vett{q},t)p(\vett{k},\vett{k}'-\vett{q})-G_c(\vett{k}+\vett{q},\vett{q},t)p(\vett{k}+\vett{q},\vett{k}')\nonumber\\
&-iG_s(\vett{k}+\vett{q},\vett{q},t)n_e(\vett{k}+\vett{q},\vett{k}')+iG_s(\vett{k}',\vett{q},t)\left[\delta_{k,k'-q}-n_h(\vett{k}'-\vett{q},\vett{k})\right]\Bigg\}
\end{align}
\eseq
\end{widetext}
where $E(\vett{k})$ [$E(\vett{k}')$] are the energy eigenvalues of valence and conduction bands in state $\vett{k}$ ($\vett{k}'$), and we have defined
\bseq
\begin{align}
G_c(\vett{k},\vett{q},t)&=\left|\Omega(\vett{k},\vett{q},t)\right|\cos\Phi(\vett{k},\vett{q},t),\\
G_s(\vett{k},\vett{q},t)&=\left|\Omega(\vett{k},\vett{q},t)\right|\sin\Phi(\vett{k},\vett{q},t).
\end{align}
\eseq
The equations above constitute a generalisation of the usual Dirac-Bloch equations, to account for the presence of a spatially varying electromagnetic pulse. This is the second result of our work. Notice, moreover, that written in the above form, the generalised Dirac-Bloch equations have a similar form that the semiconductor Bloch equations for an inhomogeneous electric field, as given in Ref. \citenum{refN}, with the difference, that we don't consider the effect of the Coulomb potential in the specific case of graphene. Accounting for the effect of Coulomb interactions on Dirac-Bloch equations, is of importance for certain categories of 2D materials, such as TMDs, for example, and will be the subject of a future publication.

\section{Current}\label{sec:current}
As a final step towards building the nonlinear optical response of graphene, we need to calculate the induced (nonlinear) current. To do so, we use the usual definition of current, in terms of population and polarisation 
\beq\label{eqCurrent}
\hat{\vett{j}}(t)=\sum_{k',k,q}\hat{\psi}^{\dagger}_{k'}U^{\dagger}(\vett{k}')\,\nabla_k\,h_I(\vett{k},\vett{q})\,U(\vett{k}-\vett{q})\hat{\psi}_{k-q},
\eeq
where, again, we work in the interaction picture, so that the $k$-dependent interaction Hamiltonian above, in the band representation, is derived directly from Eq. \eqref{eq37bis}, i.e.,
\barr\label{eq58}
h_I(\vett{k},\vett{q})&=&\hbar\,\boldsymbol\sigma\cdot\boldsymbol\Omega,
\earr
where $\boldsymbol\Omega=\Omega_x(\vett{k},\vett{q},t)\uvett{x}+\Omega_y(\vett{k},\vett{q},t)\uvett{y}$, $\boldsymbol\sigma=\sigma_x\uvett{x}+\sigma_y\uvett{y}$,
 $\hat{\psi}_k=(\annk{e}{k}\hspace{1mm} \,\crek{h}{-k})^T$, and $U(\vett{k})$ being the matrix whose columns are the eigenstates of $\hat{H}_0$ calculated before. If we then define the following $2\times 2$
 matrices
\barr
M_x&=&U^{\dagger}\nabla_kh_I\,U\Big|_x, \\ \nonumber
M_y&=&U^{\dagger}\nabla_kh_I\,U\Big|_y,
\earr
and call $m^{x,y}_{ij}(\vett{k},\vett{k}',\vett{q})$ their matrix elements, whose explicit expressions are given as follows
\bseq
\begin{align}
m_{11}^x&=|\Omega_0(\vett{q},t)|\Big[-E_x\sin2\varphi(\vett{k}-\vett{q})\cos\delta_x(\vett{k}',\vett{k}-\vett{q})\nonumber\\
&+E_y\cos2\varphi(\vett{k}-\vett{q})\cos\delta_y(\vett{k}',\vett{k}-\vett{q})\Big],\\
m_{12}^x &=-i|\Omega_0(\vett{q},t)|\Big[E_x\sin2\varphi(\vett{k}-\vett{q})\sin\delta_x(\vett{k}',\vett{k}-\vett{q})\nonumber\\
&-E_y\cos2\varphi(\vett{k}-\vett{q})\sin\delta_y(\vett{k}',\vett{k}-\vett{q})\Big],\\
m_{11}^y&=|\Omega_0(\vett{q},t)|\Big[E_x\cos2\varphi(\vett{k}-\vett{q})\cos\delta_x(\vett{k}',\vett{k}-\vett{q})\nonumber\\
&+E_y\sin2\varphi(\vett{k}-\vett{q})\cos\delta_y(\vett{k}',\vett{k}-\vett{q})\Big],\\
m_{12}^y &=i|\Omega_0(\vett{q},t)|\Big[E_x\cos2\varphi(\vett{k}-\vett{q})\sin\delta_x(\vett{k}',\vett{k}-\vett{q})\nonumber\\
&+E_y\sin2\varphi(\vett{k}-\vett{q})\sin\delta_y(\vett{k}',\vett{k}-\vett{q})\Big],
\end{align}
\eseq
where $\delta_k(\vett{k}',\vett{k}-\vett{q})=\alpha_k+\frac{\varphi(\vett{k}')+\varphi(\vett{k}-\vett{q})}{2}$, the current operator $\hat{\vett{j}}=\hat{j}_x\uvett{x}+\hat{j}_y\uvett{y}$ in Eq. \eqref{eqCurrent} can be then written as follows
\bseq
\barr
\hat{j}_x(t)&=&\sum_{k,q}(m_{11}^x\crek{e}{k'}\annk{e}{k-q}+m_{12}^x\crek{e}{k'}\crek{h}{-k+q} \\ \nonumber
&+&m_{21}^x\annk{h}{-k'}\annk{e}{k-q}+m_{22}^x\annk{h}{-k'}\crek{h}{-k+q}),
\earr
\barr
\hat{j}_y(t)&=&\sum_{k,q}(m_{11}^y\crek{e}{k'}\annk{e}{k-q}+m_{12}^y\crek{e}{k'}\crek{h}{-k+q} \\ \nonumber
&+&m_{21}^y\annk{h}{-k'}\annk{e}{k-q}+m_{22}^y\annk{h}{-k'}\crek{h}{-k+q}).
\earr
\eseq
Taking the expectation values we finally get
\begin{widetext}
\bseq \label{current}
\begin{align}
j_x(t)&=\sum_{k',k,q}\Bigg\{-m_{11}^x(\vett{k}',\vett{k},\vett{q},t)\left[1-n_e(\vett{k}',\vett{k}-\vett{q})-n_h(\vett{k}',\vett{k}-\vett{q})\right]
-2\operatorname{Im}\Big\{\tilde{m}_{12}^x(\vett{k}',\vett{k},\vett{q},t)\, p(\vett{k}',\vett{k}-\vett{q})\Big\}\Bigg\}, \\
j_y(t)&=\sum_{k',k,q}\Bigg\{-m_{11}^y(\vett{k}',\vett{k},\vett{q},t)\left[1-n_e(\vett{k}',\vett{k}-\vett{q})-n_h(\vett{k}',\vett{k}-\vett{q})\right]
+2\operatorname{Im}\Big\{\tilde{m}_{12}^y(\vett{k}',\vett{k},\vett{q},t)\, p(\vett{k}',\vett{k}-\vett{q})\Big\}\Bigg\},
\end{align}
\eseq
\end{widetext}
where we have used the fact that $m_{11}^{x,y}=-m_{22}^{x,y}$ and $m_{12}^{x,y}=i\tilde{m}_{12}^{x,y}=-i(\tilde{m}_{21}^{x,y})^*$.

Notice, that the current above has the same formal expression that the one for the homogeneous case, once the replacements $n_{e,h}(\vett{k},\vett{k}')\rightarrow n_{e,h}(\vett{k})$, $p(\vett{k},\vett{k}')\rightarrow p(\vett{k})$, and $m_{ij}^{x,y}(\vett{k}',\vett{k},\vett{q},t)\rightarrow\,m_{ij}^{x,y}(\vett{k},\vett{k},0,t)$ are made.

\section{Nonlinear optical response}\label{sec:nonOptRes}
\begin{figure}[!t]
    \centering
    \includegraphics[width=\linewidth]{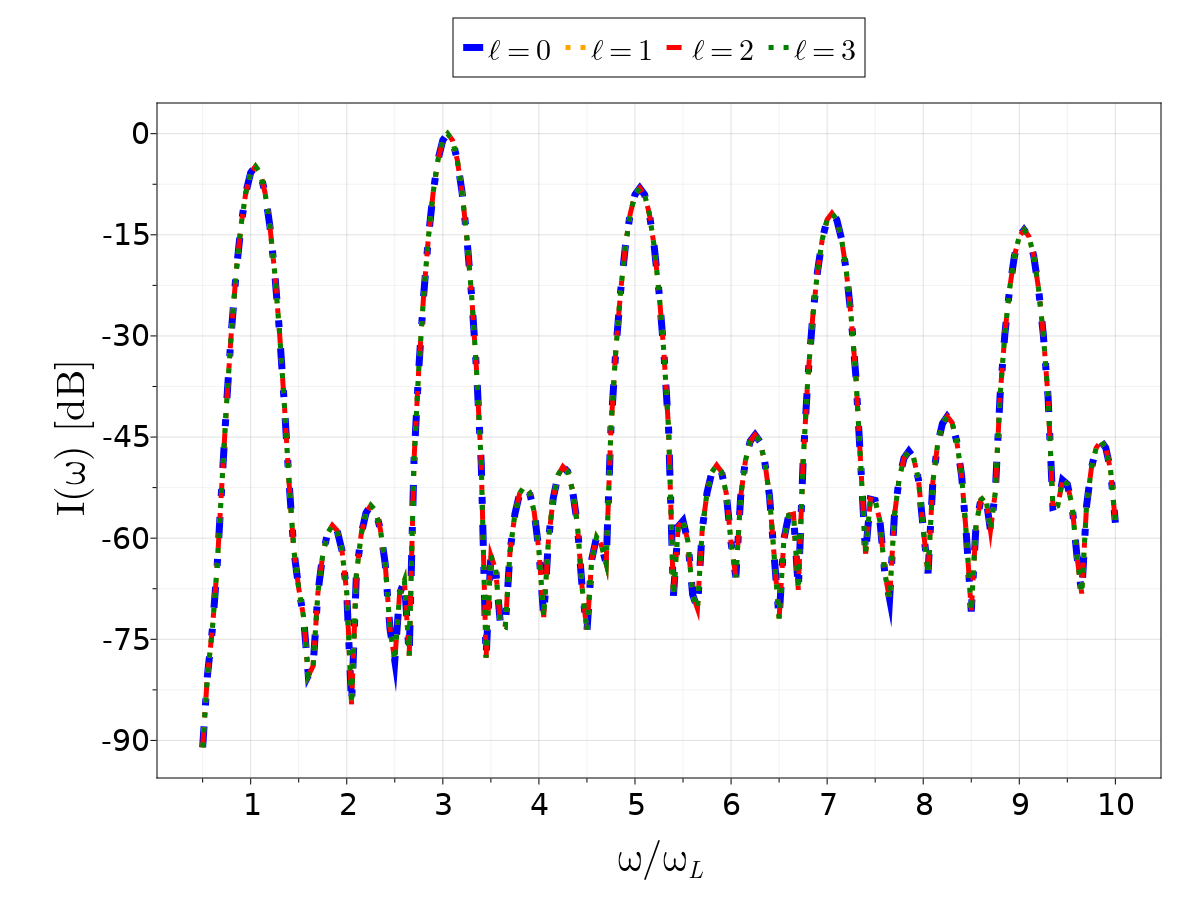}
    \caption{Nonlinear Optical Response $I(\omega)$ of graphene, as a function of the different harmonics of the carrier frequency $\omega_L$ of the impinging Laguerre-Gaussian electromagnetic pulse, for different values of the OAM parameter $\ell$, namely $\ell=0$ (dotted, blue line), $\ell=1$ (dashed, orange line), $\ell=2$ (dotted, red line), and $\ell=3$ (dashed, green line). To generate this figure, the following parameters have been used: $E_0=10^7$ V/m, $\tau=50$ fs, $\omega_L=1$ THz. Moreover, $z=0$ and $w_0=1$ (corresponding to scaled, transverse coordinates) have been assumed. }
    \label{fig:2k_laguerre_OAM}
\end{figure}
To check the validity of our approach, in this section we calculate the nonlinear optical response of graphene, when excited by a structured light pulse. To this aim, and without loss of generality, we assume the impinging pulse to be polarised along the $x$-direction, and we choose its temporal profile of the pulse to be Gaussian, while its spatial intensity distribution matches that of a Laguerre-Gaussian (LG) beam,i.e.,
\beq
\vett{E}(\vett{r},t)=E_0e^{-\frac{(t-t_0)^2}{\tau^2}-\frac{R^2}{w^2(z)}}u(\vett{R},z)\cos\omega_Lt\,\uvett{x},
\eeq
where $\tau$ is the pulse duration, $t_0$ an arbitrary offset, $\vett{R}=\{R,\phi\}$, and
\beq
u(x,y,z)=\frac{C_{\ell,p}}{w(z)}\left(\frac{R\sqrt{2}}{w(z)}\right)^{|\ell|}\text{L}_p^{|\ell|}\left(\frac{2R^2}{w^2(z)}\right)e^{i\ell\phi},
\eeq
with $w(z)=w_0\sqrt{1+(z/z_R)^2}$ is the Gaussian beam waist (with $z_R$ being the Rayleigh range of the LG-beam), $C_{\ell,p}=w_0\sqrt{2p!/[\pi(p+|\ell|)!]}$, and $\text{L}_p^{|\ell|}(x)$ are the associated Laguerre polynomials \cite{nist}, and $\ell$ represents the OAM index, while $p$ represents the radial index of the LG-beam.
\begin{figure}[!t]
    \centering
    \includegraphics[width=\linewidth]{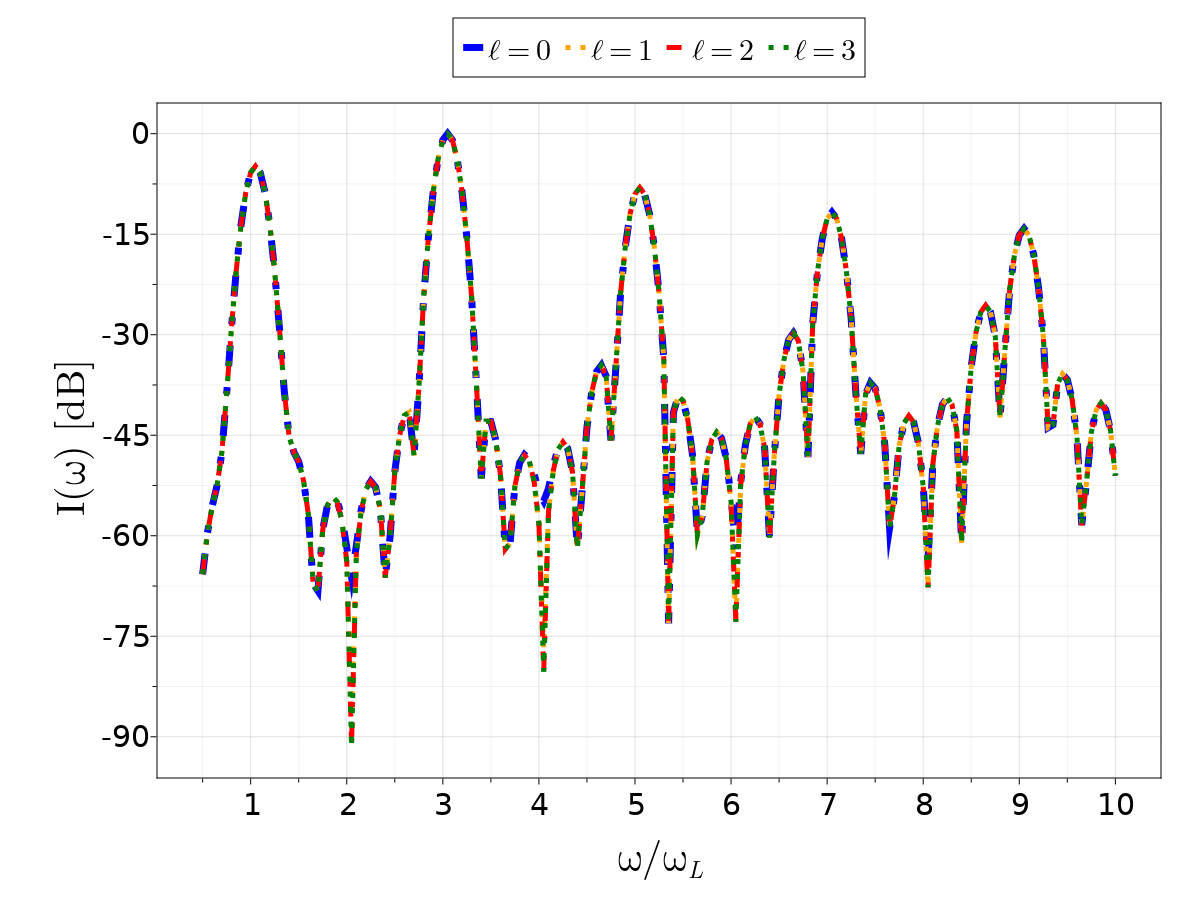}
    \caption{Nonlinear Optical Response $I(\omega)$ of graphene as a function of the different harmonics of the carrier frequency $\omega_L$ of the impinging Laguerre-Gaussian electromagnetic pulse, for different values of the OAM parameter $\ell$, namely $\ell=0$ (dashed, blue line), $\ell=1$ (dotted, orange line), $\ell=2$, (dashed, red line), and $\ell=3$ (dotted, green line). The parameters for this simulations are the same as those used for generating Fig. \ref{fig:2k_laguerre_OAM}. Moreover, this figure has been calculated under the assumption $\vett{k}=\vett{k}'$. Notice, how this assumption essentially removed any dependence on the transverse intensity distribution, levelling all the different OAM modes to the same result obtained for $\ell=0$. }
    \label{fig:1k_laguerre_OAM}
\end{figure}
We then solve Eqs. \eqref{DiracBloch} for different values of the OAM parameter $\ell$ (and set $p=0$, for simplicity), and then use Eqs. \eqref{current} to compute the correspondent nonlinear current $\vett{j}(t)$ by integrating over the transverse shape of the pulse, and we then calculate the nonlinear signal as $I(\omega) \thicksim | \omega\, \mathbf{J}(\omega) |^2$, where $\mathbf{J}(\omega)$ is the Fourier transform of the nonlinear current.

The resulting nonlinear spectrum for different values of the OAM parameter $\ell$ is shown in Fig. \ref{fig:2k_laguerre_OAM}. At a first glance, this spectrum shares the same characteristics of the usual harmonic spectrum obtained for graphene, i.e., the odd harmonics are predominant in magnitude than the even one, since the even ones are suppressed because of symmetry reasons \cite{katsnelson}. 

If we take a closer look at the even harmonics, however, we can see that they are not completely suppressed, as it happens for the case of graphene [see Fig. \ref{fig:1k_laguerre_OAM}]. This is an indirect result of the fact, that the intensity of the impinging pulse was not uniform overall the graphene sample, but had the characteristic doughnut shape of LG modes. The emergence of even harmonics in the nonlinear spectrum of a monolayer of graphene has been reported before \cite{fabio1} and it has been explained in terms of a pulse-induced oscillation of the Dirac cone, that ultimately amplifies even harmonics. In the context of this manuscript, we observe the same effect, but triggered by the non-uniform spatial intensity distribution of the pulse, rather than its temporal features. To corroborate this statement, we can compare the results in Fig. \ref{fig:2k_laguerre_OAM}, obtained by solving the full Dirac-Bloch system [Eqs. \eqref{DiracBloch}], with Fig. \ref{fig:1k_laguerre_OAM}, which essentially corresponds to the uniform intensity case (see below). As it can be seen, while the even harmonics are completely suppressed in the uniform illumination case, for the case of structured illumination reported in Fig. \ref{fig:2k_laguerre_OAM}, the actual non-uniform spatial intensity distribution has the overall effect to slightly break graphene's centrosymmetry, thus allowing the generation of even harmonics.
\subsection{A Remark on the numerical model}
It is worth noticing, that to obtain the result shown in Fig. \ref{fig:2k_laguerre_OAM}, we have used the expression of the current given by Eq. \eqref{eqCurrent}, which is a time-dependent only quantity, since it is the result of the integration, over the 4-dimensional reciprocal space $\{\vett{k},\vett{k}'\}$, of the $k$-dependent current density. If we want to have information on the spatial variation of the current or, analogously, of the nonlinear spectrum, we would instead need to compute the spatial Fourier transform of the current density, instead of simply integrating. By doing so, we will obtain the following, nonlocal, expression for the spatially-dependent nonlinear current
\bseq\label{currentNonlocal}
\begin{align}
j_x(\vett{R},\vett{R}',t)&=\sum_{k,k',q}\mathcal{J}_x(\vett{k},\vett{k}',\vett{q},t)e^{i(\vett{k}\cdot\vett{R}+\vett{k}'\cdot\vett{R}')},\\
j_y(\vett{R},\vett{R}',t)&=\sum_{k,k',q}\mathcal{J}_y(\vett{k},\vett{k}',\vett{q},t)e^{i(\vett{k}\cdot\vett{R}+\vett{k}'\cdot\vett{R}')},
\end{align}
\eseq
where $\mathcal{J}_{x,y}(\vett{k},\vett{k}',t)$ are the expressions inside the curly brackets in Eqs. \eqref{current}. Calculating the Fourier transforms above, however, is computationally very demanding, since the current densities $\mathcal{J}_{x,y}(\vett{k},\vett{k}',t)$ are defined on an 8-dimensional grid (since they are functions of three independent $k$-vectors, namely $\vett{k}$, \vett{k}', and $\vett{q}$), which evolves in time. To efficiently perform these calculations, one would probably need to solve Eqs. \eqref{DiracBloch} in a more efficient way, optimising the code and using, for example, a supercomputer, and then be very careful in defining a big enough grid in both $\vett{k}$ and $\vett{k}'$ to ensure enough resolution for the Fourier transform. Doing this, however, would require a considerable amount of time and it is outside the scope of this work.

To partly solve this issue, however, we suggest a simplification of the problem, which results in significantly reducing the computational complexity of the problem, and could be employed to make qualitative predictions on the spatial distribution of the current. To do that, we restrict the 4-dimensional $k$-space to an effective 2-dimensional one by setting $\vett{k}=\vett{k}'$ in Eqs. \eqref{DiracBloch}. This, essentially, means, that we restrict the domain of definition of $\vett{k}'$ to a point, rather than a plane. To test the validity of this approximation, in Fig. \ref{fig:1k_laguerre_OAM} we again show the harmonic spectrum and notice, that we get a pretty similar, at least qualitatively, result, than the one obtained in fig. \ref{fig:2k_laguerre_OAM} using the full model. Notice, in particular, how setting $\vett{k}=\vett{k}'$ essentially erases any information about the even harmonics. This, again, is not surprising, since the limit $\vett{k}=\vett{k}'$ can be interpreted, to some extent, as the homogeneous field limit.

Under this approximation, however, we would be able to look at the qualitative spatial evolution of the current, as now the components of the spatially dependent nonlinear current are much easier to calculate, as their explicit expressions now only amount to a regular spatial Fourier transform, i.e., 
\bseq
\begin{align}
j_x(\vett{R},t)&=\sum_{k,q}\mathcal{J}_x(\vett{k},\vett{q},t)e^{i\vett{k}\cdot\vett{R}},\\
j_y(\vett{R},t)&=\sum_{k,q}\mathcal{J}_y(\vett{k},\vett{q},t)e^{i\vett{k}\cdot\vett{R}}.
\end{align}
\eseq
Despite the equations above are greatly simplified, in terms of computational challenge, with respect to Eqs. \eqref{currentNonlocal}, they still need quite a high resolution on the $k$-space grid, in order to be able to generate meaningful images in direct space. Although the implementation of such optimised code is out of the scope of this manuscript, we intend to work on it for a future publication, where we aim at showing the different current dynamics corresponding to different spatial profiles of the impinging pulse.

\section{Conclusion}\label{sec:conc}
In this work, we introduced a general framework to study the optical response of graphene under the action of spatially-varying electromagnetic pulses, i.e., when graphene interacts with structured light. In particular, we have provided detailed derivation of the spatially-varying interaction Hamiltonian, as well as the two-body population and polarisation, constituting the building blocks for the Dirac-Bloch equations. We have then calculated the expression of the induced nonlinear current and used it to test our model against known results, by calculating the nonlinear spectrum for the case of an impinging Laguerre-Gaussian optical pulse. Surprisingly, although the overall structure of the nonlinear spectrum is not changed by a spatially varying electromagnetic field, we have reported how a spatially inhomogeneous pulse can locally break the centrosymmetry of graphene and allow the generation of even harmonics. Finally, we have shown how to obtain qualitative results to study the spatial evolution of the current density. 

Although in this work we have focused our attention to graphene, our results can be easily adapted to any 2D material and provide a general framework to include the effects of structured light in both the linear and nonlinear optical response of such materials. 

\section*{Acknowledgements}
We acknowledge the financial support of the Academy of Finland Flagship Programme, Photonics research and innovation (PREIN), decision 320165. Y.T. also acknowledges support from the Finnish Cultural Foundation, decision 00221008.

\bibliography{Bibliography}.

\begin{thebibliography}{37}%
\makeatletter
\providecommand \@ifxundefined [1]{%
 \@ifx{#1\undefined}
}%
\providecommand \@ifnum [1]{%
 \ifnum #1\expandafter \@firstoftwo
 \else \expandafter \@secondoftwo
 \fi
}%
\providecommand \@ifx [1]{%
 \ifx #1\expandafter \@firstoftwo
 \else \expandafter \@secondoftwo
 \fi
}%
\providecommand \natexlab [1]{#1}%
\providecommand \enquote  [1]{``#1''}%
\providecommand \bibnamefont  [1]{#1}%
\providecommand \bibfnamefont [1]{#1}%
\providecommand \citenamefont [1]{#1}%
\providecommand \href@noop [0]{\@secondoftwo}%
\providecommand \href [0]{\begingroup \@sanitize@url \@href}%
\providecommand \@href[1]{\@@startlink{#1}\@@href}%
\providecommand \@@href[1]{\endgroup#1\@@endlink}%
\providecommand \@sanitize@url [0]{\catcode `\\12\catcode `\$12\catcode
  `\&12\catcode `\#12\catcode `\^12\catcode `\_12\catcode `\%12\relax}%
\providecommand \@@startlink[1]{}%
\providecommand \@@endlink[0]{}%
\providecommand \url  [0]{\begingroup\@sanitize@url \@url }%
\providecommand \@url [1]{\endgroup\@href {#1}{\urlprefix }}%
\providecommand \urlprefix  [0]{URL }%
\providecommand \Eprint [0]{\href }%
\providecommand \doibase [0]{http://dx.doi.org/}%
\providecommand \selectlanguage [0]{\@gobble}%
\providecommand \bibinfo  [0]{\@secondoftwo}%
\providecommand \bibfield  [0]{\@secondoftwo}%
\providecommand \translation [1]{[#1]}%
\providecommand \BibitemOpen [0]{}%
\providecommand \bibitemStop [0]{}%
\providecommand \bibitemNoStop [0]{.\EOS\space}%
\providecommand \EOS [0]{\spacefactor3000\relax}%
\providecommand \BibitemShut  [1]{\csname bibitem#1\endcsname}%
\let\auto@bib@innerbib\@empty
\bibitem [{\citenamefont {Novoselov}\ \emph {et~al.}(2004)\citenamefont
  {Novoselov}, \citenamefont {Geim}, \citenamefont {Morozov}, \citenamefont
  {Jiang}, \citenamefont {Zhang}, \citenamefont {Dubonos}, \citenamefont
  {Grigorieva},\ and\ \citenamefont {Firsov}}]{novoselov_electric_2004}%
  \BibitemOpen
  \bibfield  {author} {\bibinfo {author} {\bibfnamefont {K.~S.}\ \bibnamefont
  {Novoselov}}, \bibinfo {author} {\bibfnamefont {A.~K.}\ \bibnamefont {Geim}},
  \bibinfo {author} {\bibfnamefont {S.~V.}\ \bibnamefont {Morozov}}, \bibinfo
  {author} {\bibfnamefont {D.}~\bibnamefont {Jiang}}, \bibinfo {author}
  {\bibfnamefont {Y.}~\bibnamefont {Zhang}}, \bibinfo {author} {\bibfnamefont
  {S.~V.}\ \bibnamefont {Dubonos}}, \bibinfo {author} {\bibfnamefont {I.~V.}\
  \bibnamefont {Grigorieva}}, \ and\ \bibinfo {author} {\bibfnamefont {A.~A.}\
  \bibnamefont {Firsov}},\ }\href {\doibase 10.1126/science.1102896} {\bibfield
   {journal} {\bibinfo  {journal} {Science}\ }\textbf {\bibinfo {volume}
  {306}},\ \bibinfo {pages} {666} (\bibinfo {year} {2004})}\BibitemShut
  {NoStop}%
\bibitem [{\citenamefont {Katsnelson}(2006)}]{refCond1}%
  \BibitemOpen
  \bibfield  {author} {\bibinfo {author} {\bibfnamefont {M.}~\bibnamefont
  {Katsnelson}},\ }\href@noop {} {\bibfield  {journal} {\bibinfo  {journal}
  {The European Physical Journal B-Condensed Matter and Complex Systems}\
  }\textbf {\bibinfo {volume} {51}},\ \bibinfo {pages} {157} (\bibinfo {year}
  {2006})}\BibitemShut {NoStop}%
\bibitem [{\citenamefont {Nair}\ \emph {et~al.}()\citenamefont {Nair},
  \citenamefont {Blake}, \citenamefont {Grigorenko}, \citenamefont {Novoselov},
  \citenamefont {Booth}, \citenamefont {Stauber}, \citenamefont {Peres},\ and\
  \citenamefont {Geim}}]{univAbs1}%
  \BibitemOpen
  \bibfield  {author} {\bibinfo {author} {\bibfnamefont {R.~R.}\ \bibnamefont
  {Nair}}, \bibinfo {author} {\bibfnamefont {P.}~\bibnamefont {Blake}},
  \bibinfo {author} {\bibfnamefont {A.~N.}\ \bibnamefont {Grigorenko}},
  \bibinfo {author} {\bibfnamefont {K.~S.}\ \bibnamefont {Novoselov}}, \bibinfo
  {author} {\bibfnamefont {T.~J.}\ \bibnamefont {Booth}}, \bibinfo {author}
  {\bibfnamefont {T.}~\bibnamefont {Stauber}}, \bibinfo {author} {\bibfnamefont
  {N.~M.~R.}\ \bibnamefont {Peres}}, \ and\ \bibinfo {author} {\bibfnamefont
  {A.~K.}\ \bibnamefont {Geim}},\ }\href {\doibase 10.1126/science.1156965} {\
  \textbf {\bibinfo {volume} {320}},\ \bibinfo {pages} {1308}}\BibitemShut
  {NoStop}%
\bibitem [{\citenamefont {Novoselov}\ \emph {et~al.}(2005)\citenamefont
  {Novoselov}, \citenamefont {Geim}, \citenamefont {Morozov}, \citenamefont
  {Jiang}, \citenamefont {Katsnelson}, \citenamefont {Grigorieva},
  \citenamefont {Dubonos},\ and\ \citenamefont {Firsov}}]{novoselov_qhe}%
  \BibitemOpen
  \bibfield  {author} {\bibinfo {author} {\bibfnamefont {K.~S.}\ \bibnamefont
  {Novoselov}}, \bibinfo {author} {\bibfnamefont {A.~K.}\ \bibnamefont {Geim}},
  \bibinfo {author} {\bibfnamefont {S.~V.}\ \bibnamefont {Morozov}}, \bibinfo
  {author} {\bibfnamefont {D.}~\bibnamefont {Jiang}}, \bibinfo {author}
  {\bibfnamefont {M.~I.}\ \bibnamefont {Katsnelson}}, \bibinfo {author}
  {\bibfnamefont {I.~V.}\ \bibnamefont {Grigorieva}}, \bibinfo {author}
  {\bibfnamefont {S.~V.}\ \bibnamefont {Dubonos}}, \ and\ \bibinfo {author}
  {\bibfnamefont {A.~A.}\ \bibnamefont {Firsov}},\ }\href {\doibase
  10.1038/nature04233} {\bibfield  {journal} {\bibinfo  {journal} {Nature}\
  }\textbf {\bibinfo {volume} {438}},\ \bibinfo {pages} {197} (\bibinfo {year}
  {2005})}\BibitemShut {NoStop}%
\bibitem [{zha()}]{zhang_qhe}%
  \BibitemOpen
  \href@noop {} {\ }\BibitemShut {NoStop}%
\bibitem [{\citenamefont {Wright}\ \emph {et~al.}(2009)\citenamefont {Wright},
  \citenamefont {Xu}, \citenamefont {Cao},\ and\ \citenamefont
  {Zhang}}]{wright_strong_2009}%
  \BibitemOpen
  \bibfield  {author} {\bibinfo {author} {\bibfnamefont {A.~R.}\ \bibnamefont
  {Wright}}, \bibinfo {author} {\bibfnamefont {X.~G.}\ \bibnamefont {Xu}},
  \bibinfo {author} {\bibfnamefont {J.~C.}\ \bibnamefont {Cao}}, \ and\
  \bibinfo {author} {\bibfnamefont {C.}~\bibnamefont {Zhang}},\ }\href
  {\doibase 10.1063/1.3205115} {\bibfield  {journal} {\bibinfo  {journal}
  {Applied Physics Letters}\ }\textbf {\bibinfo {volume} {95}},\ \bibinfo
  {pages} {072101} (\bibinfo {year} {2009})}\BibitemShut {NoStop}%
\bibitem [{\citenamefont {Hafez}\ \emph {et~al.}(2020)\citenamefont {Hafez},
  \citenamefont {Kovalev}, \citenamefont {Tielrooij}, \citenamefont {Bonn},
  \citenamefont {Gensch},\ and\ \citenamefont
  {Turchinovich}}]{hafez_terahertz_2020}%
  \BibitemOpen
  \bibfield  {author} {\bibinfo {author} {\bibfnamefont {H.~A.}\ \bibnamefont
  {Hafez}}, \bibinfo {author} {\bibfnamefont {S.}~\bibnamefont {Kovalev}},
  \bibinfo {author} {\bibfnamefont {K.}~\bibnamefont {Tielrooij}}, \bibinfo
  {author} {\bibfnamefont {M.}~\bibnamefont {Bonn}}, \bibinfo {author}
  {\bibfnamefont {M.}~\bibnamefont {Gensch}}, \ and\ \bibinfo {author}
  {\bibfnamefont {D.}~\bibnamefont {Turchinovich}},\ }\href {\doibase
  10.1002/adom.201900771} {\bibfield  {journal} {\bibinfo  {journal} {Advanced
  Optical Materials}\ }\textbf {\bibinfo {volume} {8}},\ \bibinfo {pages}
  {1900771} (\bibinfo {year} {2020})}\BibitemShut {NoStop}%
\bibitem [{\citenamefont {Manzeli}\ \emph {et~al.}(2017)\citenamefont
  {Manzeli}, \citenamefont {Ovchinnikov}, \citenamefont {Pasquier},
  \citenamefont {Yazyev},\ and\ \citenamefont {Kis}}]{manzeli_2d_2017}%
  \BibitemOpen
  \bibfield  {author} {\bibinfo {author} {\bibfnamefont {S.}~\bibnamefont
  {Manzeli}}, \bibinfo {author} {\bibfnamefont {D.}~\bibnamefont
  {Ovchinnikov}}, \bibinfo {author} {\bibfnamefont {D.}~\bibnamefont
  {Pasquier}}, \bibinfo {author} {\bibfnamefont {O.~V.}\ \bibnamefont
  {Yazyev}}, \ and\ \bibinfo {author} {\bibfnamefont {A.}~\bibnamefont {Kis}},\
  }\href {\doibase 10.1038/natrevmats.2017.33} {\bibfield  {journal} {\bibinfo
  {journal} {Nature Reviews Materials}\ }\textbf {\bibinfo {volume} {2}},\
  \bibinfo {pages} {17033} (\bibinfo {year} {2017})}\BibitemShut {NoStop}%
\bibitem [{\citenamefont {Autere}\ \emph {et~al.}(2018)\citenamefont {Autere},
  \citenamefont {Jussila}, \citenamefont {Dai}, \citenamefont {Wang},
  \citenamefont {Lipsanen},\ and\ \citenamefont {Sun}}]{autere_nonlinear_2018}%
  \BibitemOpen
  \bibfield  {author} {\bibinfo {author} {\bibfnamefont {A.}~\bibnamefont
  {Autere}}, \bibinfo {author} {\bibfnamefont {H.}~\bibnamefont {Jussila}},
  \bibinfo {author} {\bibfnamefont {Y.}~\bibnamefont {Dai}}, \bibinfo {author}
  {\bibfnamefont {Y.}~\bibnamefont {Wang}}, \bibinfo {author} {\bibfnamefont
  {H.}~\bibnamefont {Lipsanen}}, \ and\ \bibinfo {author} {\bibfnamefont
  {Z.}~\bibnamefont {Sun}},\ }\href {\doibase 10.1002/adma.201705963}
  {\bibfield  {journal} {\bibinfo  {journal} {Advanced Materials}\ }\textbf
  {\bibinfo {volume} {30}},\ \bibinfo {pages} {1705963} (\bibinfo {year}
  {2018})}\BibitemShut {NoStop}%
\bibitem [{\citenamefont {Neto}\ \emph {et~al.}(2009)\citenamefont {Neto},
  \citenamefont {Guinea}, \citenamefont {Peres}, \citenamefont {Novoselov},\
  and\ \citenamefont {Geim}}]{neto2009electronic}%
  \BibitemOpen
  \bibfield  {author} {\bibinfo {author} {\bibfnamefont {A.~C.}\ \bibnamefont
  {Neto}}, \bibinfo {author} {\bibfnamefont {F.}~\bibnamefont {Guinea}},
  \bibinfo {author} {\bibfnamefont {N.~M.}\ \bibnamefont {Peres}}, \bibinfo
  {author} {\bibfnamefont {K.~S.}\ \bibnamefont {Novoselov}}, \ and\ \bibinfo
  {author} {\bibfnamefont {A.~K.}\ \bibnamefont {Geim}},\ }\href@noop {}
  {\bibfield  {journal} {\bibinfo  {journal} {Reviews of modern physics}\
  }\textbf {\bibinfo {volume} {81}},\ \bibinfo {pages} {109} (\bibinfo {year}
  {2009})}\BibitemShut {NoStop}%
\bibitem [{\citenamefont {Caldwell}\ \emph {et~al.}(2019)\citenamefont
  {Caldwell}, \citenamefont {Aharonovich}, \citenamefont {Cassabois},
  \citenamefont {Edgar}, \citenamefont {Gil},\ and\ \citenamefont
  {Basov}}]{caldwell_photonics_2019}%
  \BibitemOpen
  \bibfield  {author} {\bibinfo {author} {\bibfnamefont {J.~D.}\ \bibnamefont
  {Caldwell}}, \bibinfo {author} {\bibfnamefont {I.}~\bibnamefont
  {Aharonovich}}, \bibinfo {author} {\bibfnamefont {G.}~\bibnamefont
  {Cassabois}}, \bibinfo {author} {\bibfnamefont {J.~H.}\ \bibnamefont
  {Edgar}}, \bibinfo {author} {\bibfnamefont {B.}~\bibnamefont {Gil}}, \ and\
  \bibinfo {author} {\bibfnamefont {D.~N.}\ \bibnamefont {Basov}},\ }\href
  {\doibase 10.1038/s41578-019-0124-1} {\bibfield  {journal} {\bibinfo
  {journal} {Nature Reviews Materials}\ }\textbf {\bibinfo {volume} {4}},\
  \bibinfo {pages} {552} (\bibinfo {year} {2019})}\BibitemShut {NoStop}%
\bibitem [{\citenamefont {Geim}\ and\ \citenamefont {Novoselov}()}]{blackPh}%
  \BibitemOpen
  \bibfield  {author} {\bibinfo {author} {\bibfnamefont {A.~K.}\ \bibnamefont
  {Geim}}\ and\ \bibinfo {author} {\bibfnamefont {K.~S.}\ \bibnamefont
  {Novoselov}},\ }\href {\doibase 10.1038/nmat1849} {\ \textbf {\bibinfo
  {volume} {6}},\ \bibinfo {pages} {183}}\BibitemShut {NoStop}%
\bibitem [{\citenamefont {Zeng}\ \emph {et~al.}(2013)\citenamefont {Zeng},
  \citenamefont {Liu}, \citenamefont {Dai}, \citenamefont {Yan}, \citenamefont
  {Zhu}, \citenamefont {He}, \citenamefont {Xie}, \citenamefont {Xu},
  \citenamefont {Chen}, \citenamefont {Yao} \emph {et~al.}}]{zeng2013optical}%
  \BibitemOpen
  \bibfield  {author} {\bibinfo {author} {\bibfnamefont {H.}~\bibnamefont
  {Zeng}}, \bibinfo {author} {\bibfnamefont {G.-B.}\ \bibnamefont {Liu}},
  \bibinfo {author} {\bibfnamefont {J.}~\bibnamefont {Dai}}, \bibinfo {author}
  {\bibfnamefont {Y.}~\bibnamefont {Yan}}, \bibinfo {author} {\bibfnamefont
  {B.}~\bibnamefont {Zhu}}, \bibinfo {author} {\bibfnamefont {R.}~\bibnamefont
  {He}}, \bibinfo {author} {\bibfnamefont {L.}~\bibnamefont {Xie}}, \bibinfo
  {author} {\bibfnamefont {S.}~\bibnamefont {Xu}}, \bibinfo {author}
  {\bibfnamefont {X.}~\bibnamefont {Chen}}, \bibinfo {author} {\bibfnamefont
  {W.}~\bibnamefont {Yao}},  \emph {et~al.},\ }\href@noop {} {\bibfield
  {journal} {\bibinfo  {journal} {Scientific reports}\ }\textbf {\bibinfo
  {volume} {3}},\ \bibinfo {pages} {1} (\bibinfo {year} {2013})}\BibitemShut
  {NoStop}%
\bibitem [{\citenamefont {Kumar}\ \emph {et~al.}(2013)\citenamefont {Kumar},
  \citenamefont {Kumar}, \citenamefont {Gerstenkorn}, \citenamefont {Wang},
  \citenamefont {Chiu}, \citenamefont {Smirl},\ and\ \citenamefont
  {Zhao}}]{kumar2013third}%
  \BibitemOpen
  \bibfield  {author} {\bibinfo {author} {\bibfnamefont {N.}~\bibnamefont
  {Kumar}}, \bibinfo {author} {\bibfnamefont {J.}~\bibnamefont {Kumar}},
  \bibinfo {author} {\bibfnamefont {C.}~\bibnamefont {Gerstenkorn}}, \bibinfo
  {author} {\bibfnamefont {R.}~\bibnamefont {Wang}}, \bibinfo {author}
  {\bibfnamefont {H.-Y.}\ \bibnamefont {Chiu}}, \bibinfo {author}
  {\bibfnamefont {A.~L.}\ \bibnamefont {Smirl}}, \ and\ \bibinfo {author}
  {\bibfnamefont {H.}~\bibnamefont {Zhao}},\ }\href@noop {} {\bibfield
  {journal} {\bibinfo  {journal} {Physical Review B}\ }\textbf {\bibinfo
  {volume} {87}},\ \bibinfo {pages} {121406} (\bibinfo {year}
  {2013})}\BibitemShut {NoStop}%
\bibitem [{\citenamefont {Hendry}\ \emph {et~al.}(2010)\citenamefont {Hendry},
  \citenamefont {Hale}, \citenamefont {Moger}, \citenamefont {Savchenko},\ and\
  \citenamefont {Mikhailov}}]{hendry_coherent_2010}%
  \BibitemOpen
  \bibfield  {author} {\bibinfo {author} {\bibfnamefont {E.}~\bibnamefont
  {Hendry}}, \bibinfo {author} {\bibfnamefont {P.~J.}\ \bibnamefont {Hale}},
  \bibinfo {author} {\bibfnamefont {J.}~\bibnamefont {Moger}}, \bibinfo
  {author} {\bibfnamefont {A.~K.}\ \bibnamefont {Savchenko}}, \ and\ \bibinfo
  {author} {\bibfnamefont {S.~A.}\ \bibnamefont {Mikhailov}},\ }\href {\doibase
  10.1103/PhysRevLett.105.097401} {\bibfield  {journal} {\bibinfo  {journal}
  {Physical Review Letters}\ }\textbf {\bibinfo {volume} {105}},\ \bibinfo
  {pages} {097401} (\bibinfo {year} {2010})}\BibitemShut {NoStop}%
\bibitem [{\citenamefont {Dong}\ \emph {et~al.}(2015)\citenamefont {Dong},
  \citenamefont {Li}, \citenamefont {Feng}, \citenamefont {Zhang},
  \citenamefont {Zhang}, \citenamefont {Chang}, \citenamefont {Fan},
  \citenamefont {Zhang},\ and\ \citenamefont {Wang}}]{dong_optical_2015}%
  \BibitemOpen
  \bibfield  {author} {\bibinfo {author} {\bibfnamefont {N.}~\bibnamefont
  {Dong}}, \bibinfo {author} {\bibfnamefont {Y.}~\bibnamefont {Li}}, \bibinfo
  {author} {\bibfnamefont {Y.}~\bibnamefont {Feng}}, \bibinfo {author}
  {\bibfnamefont {S.}~\bibnamefont {Zhang}}, \bibinfo {author} {\bibfnamefont
  {X.}~\bibnamefont {Zhang}}, \bibinfo {author} {\bibfnamefont
  {C.}~\bibnamefont {Chang}}, \bibinfo {author} {\bibfnamefont
  {J.}~\bibnamefont {Fan}}, \bibinfo {author} {\bibfnamefont {L.}~\bibnamefont
  {Zhang}}, \ and\ \bibinfo {author} {\bibfnamefont {J.}~\bibnamefont {Wang}},\
  }\href {\doibase 10.1038/srep14646} {\bibfield  {journal} {\bibinfo
  {journal} {Scientific Reports}\ }\textbf {\bibinfo {volume} {5}},\ \bibinfo
  {pages} {14646} (\bibinfo {year} {2015})}\BibitemShut {NoStop}%
\bibitem [{\citenamefont {Ornigotti}\ \emph {et~al.}(2021)\citenamefont
  {Ornigotti}, \citenamefont {Ornigotti},\ and\ \citenamefont
  {Biancalana}}]{ornigotti_strain_2021}%
  \BibitemOpen
  \bibfield  {author} {\bibinfo {author} {\bibfnamefont {M.}~\bibnamefont
  {Ornigotti}}, \bibinfo {author} {\bibfnamefont {L.}~\bibnamefont
  {Ornigotti}}, \ and\ \bibinfo {author} {\bibfnamefont {F.}~\bibnamefont
  {Biancalana}},\ }\href {\doibase 10.1063/5.0049678} {\bibfield  {journal}
  {\bibinfo  {journal} {APL Photonics}\ }\textbf {\bibinfo {volume} {6}},\
  \bibinfo {pages} {060801} (\bibinfo {year} {2021})}\BibitemShut {NoStop}%
\bibitem [{\citenamefont {Carvalho}\ \emph {et~al.}()\citenamefont {Carvalho},
  \citenamefont {Biancalana},\ and\ \citenamefont {Marini}}]{fabio1}%
  \BibitemOpen
  \bibfield  {author} {\bibinfo {author} {\bibfnamefont {D.~N.}\ \bibnamefont
  {Carvalho}}, \bibinfo {author} {\bibfnamefont {F.}~\bibnamefont
  {Biancalana}}, \ and\ \bibinfo {author} {\bibfnamefont {A.}~\bibnamefont
  {Marini}},\ }\href {\doibase 10.1515/odps-2017-0006} {\ \textbf {\bibinfo
  {volume} {3}},\ 10.1515/odps-2017-0006}\BibitemShut {NoStop}%
\bibitem [{\citenamefont {Allen}\ \emph {et~al.}(1992)\citenamefont {Allen},
  \citenamefont {Beijersbergen}, \citenamefont {Spreeuw},\ and\ \citenamefont
  {Woerdman}}]{allen_orbital_1992}%
  \BibitemOpen
  \bibfield  {author} {\bibinfo {author} {\bibfnamefont {L.}~\bibnamefont
  {Allen}}, \bibinfo {author} {\bibfnamefont {M.~W.}\ \bibnamefont
  {Beijersbergen}}, \bibinfo {author} {\bibfnamefont {R.~J.~C.}\ \bibnamefont
  {Spreeuw}}, \ and\ \bibinfo {author} {\bibfnamefont {J.~P.}\ \bibnamefont
  {Woerdman}},\ }\href {\doibase 10.1103/PhysRevA.45.8185} {\bibfield
  {journal} {\bibinfo  {journal} {Physical Review A}\ }\textbf {\bibinfo
  {volume} {45}},\ \bibinfo {pages} {8185} (\bibinfo {year}
  {1992})}\BibitemShut {NoStop}%
\bibitem [{\citenamefont {Franke-Arnold}\ \emph {et~al.}()\citenamefont
  {Franke-Arnold}, \citenamefont {Allen},\ and\ \citenamefont
  {Padgett}}]{oam1}%
  \BibitemOpen
  \bibfield  {author} {\bibinfo {author} {\bibfnamefont {S.}~\bibnamefont
  {Franke-Arnold}}, \bibinfo {author} {\bibfnamefont {L.}~\bibnamefont
  {Allen}}, \ and\ \bibinfo {author} {\bibfnamefont {M.}~\bibnamefont
  {Padgett}},\ }\href {\doibase 10.1002/lpor.200810007} {\ \textbf {\bibinfo
  {volume} {2}},\ \bibinfo {pages} {299}}\BibitemShut {NoStop}%
\bibitem [{\citenamefont {Fickler}\ \emph {et~al.}()\citenamefont {Fickler},
  \citenamefont {Lapkiewicz}, \citenamefont {Plick}, \citenamefont {Krenn},
  \citenamefont {Schaeff}, \citenamefont {Ramelow},\ and\ \citenamefont
  {Zeilinger}}]{oam2}%
  \BibitemOpen
  \bibfield  {author} {\bibinfo {author} {\bibfnamefont {R.}~\bibnamefont
  {Fickler}}, \bibinfo {author} {\bibfnamefont {R.}~\bibnamefont {Lapkiewicz}},
  \bibinfo {author} {\bibfnamefont {W.~N.}\ \bibnamefont {Plick}}, \bibinfo
  {author} {\bibfnamefont {M.}~\bibnamefont {Krenn}}, \bibinfo {author}
  {\bibfnamefont {C.}~\bibnamefont {Schaeff}}, \bibinfo {author} {\bibfnamefont
  {S.}~\bibnamefont {Ramelow}}, \ and\ \bibinfo {author} {\bibfnamefont
  {A.}~\bibnamefont {Zeilinger}},\ }\href {\doibase 10.1126/science.1227193} {\
  \textbf {\bibinfo {volume} {338}},\ \bibinfo {pages} {640}}\BibitemShut
  {NoStop}%
\bibitem [{\citenamefont {Klar}\ and\ \citenamefont {Hell}()}]{oam3}%
  \BibitemOpen
  \bibfield  {author} {\bibinfo {author} {\bibfnamefont {T.~A.}\ \bibnamefont
  {Klar}}\ and\ \bibinfo {author} {\bibfnamefont {S.~W.}\ \bibnamefont
  {Hell}},\ }\href {\doibase 10.1364/OL.24.000954} {\ \textbf {\bibinfo
  {volume} {24}},\ \bibinfo {pages} {954}}\BibitemShut {NoStop}%
\bibitem [{\citenamefont {Noyan}\ and\ \citenamefont {Kikkawa}()}]{oam4}%
  \BibitemOpen
  \bibfield  {author} {\bibinfo {author} {\bibfnamefont {M.~A.}\ \bibnamefont
  {Noyan}}\ and\ \bibinfo {author} {\bibfnamefont {J.~M.}\ \bibnamefont
  {Kikkawa}},\ }\href {\doibase 10.1063/1.4927321} {\ \textbf {\bibinfo
  {volume} {107}},\ \bibinfo {pages} {032406}}\BibitemShut {NoStop}%
\bibitem [{\citenamefont {Wang}\ \emph {et~al.}()\citenamefont {Wang},
  \citenamefont {Yang}, \citenamefont {Fazal}, \citenamefont {Ahmed},
  \citenamefont {Yan}, \citenamefont {Huang}, \citenamefont {Ren},
  \citenamefont {Yue}, \citenamefont {Dolinar}, \citenamefont {Tur},\ and\
  \citenamefont {Willner}}]{oam5}%
  \BibitemOpen
  \bibfield  {author} {\bibinfo {author} {\bibfnamefont {J.}~\bibnamefont
  {Wang}}, \bibinfo {author} {\bibfnamefont {J.-Y.}\ \bibnamefont {Yang}},
  \bibinfo {author} {\bibfnamefont {I.~M.}\ \bibnamefont {Fazal}}, \bibinfo
  {author} {\bibfnamefont {N.}~\bibnamefont {Ahmed}}, \bibinfo {author}
  {\bibfnamefont {Y.}~\bibnamefont {Yan}}, \bibinfo {author} {\bibfnamefont
  {H.}~\bibnamefont {Huang}}, \bibinfo {author} {\bibfnamefont
  {Y.}~\bibnamefont {Ren}}, \bibinfo {author} {\bibfnamefont {Y.}~\bibnamefont
  {Yue}}, \bibinfo {author} {\bibfnamefont {S.}~\bibnamefont {Dolinar}},
  \bibinfo {author} {\bibfnamefont {M.}~\bibnamefont {Tur}}, \ and\ \bibinfo
  {author} {\bibfnamefont {A.~E.}\ \bibnamefont {Willner}},\ }\href {\doibase
  10.1038/nphoton.2012.138} {\ \textbf {\bibinfo {volume} {6}},\ \bibinfo
  {pages} {488}}\BibitemShut {NoStop}%
\bibitem [{\citenamefont {Cozzolino}\ \emph {et~al.}()\citenamefont
  {Cozzolino}, \citenamefont {Bacco}, \citenamefont {Da~Lio}, \citenamefont
  {Ingerslev}, \citenamefont {Ding}, \citenamefont {Dalgaard}, \citenamefont
  {Kristensen}, \citenamefont {Galili}, \citenamefont {Rottwitt}, \citenamefont
  {Ramachandran},\ and\ \citenamefont {Oxenløwe}}]{oam6}%
  \BibitemOpen
  \bibfield  {author} {\bibinfo {author} {\bibfnamefont {D.}~\bibnamefont
  {Cozzolino}}, \bibinfo {author} {\bibfnamefont {D.}~\bibnamefont {Bacco}},
  \bibinfo {author} {\bibfnamefont {B.}~\bibnamefont {Da~Lio}}, \bibinfo
  {author} {\bibfnamefont {K.}~\bibnamefont {Ingerslev}}, \bibinfo {author}
  {\bibfnamefont {Y.}~\bibnamefont {Ding}}, \bibinfo {author} {\bibfnamefont
  {K.}~\bibnamefont {Dalgaard}}, \bibinfo {author} {\bibfnamefont
  {P.}~\bibnamefont {Kristensen}}, \bibinfo {author} {\bibfnamefont
  {M.}~\bibnamefont {Galili}}, \bibinfo {author} {\bibfnamefont
  {K.}~\bibnamefont {Rottwitt}}, \bibinfo {author} {\bibfnamefont
  {S.}~\bibnamefont {Ramachandran}}, \ and\ \bibinfo {author} {\bibfnamefont
  {L.~K.}\ \bibnamefont {Oxenløwe}},\ }\href {\doibase
  10.1103/PhysRevApplied.11.064058} {\ \textbf {\bibinfo {volume} {11}},\
  \bibinfo {pages} {064058}}\BibitemShut {NoStop}%
\bibitem [{\citenamefont {Wätzel}\ and\ \citenamefont
  {Berakdar}(2016)}]{watzel_centrifugal_2016}%
  \BibitemOpen
  \bibfield  {author} {\bibinfo {author} {\bibfnamefont {J.}~\bibnamefont
  {Wätzel}}\ and\ \bibinfo {author} {\bibfnamefont {J.}~\bibnamefont
  {Berakdar}},\ }\href {\doibase 10.1038/srep21475} {\bibfield  {journal}
  {\bibinfo  {journal} {Scientific Reports}\ }\textbf {\bibinfo {volume} {6}},\
  \bibinfo {pages} {21475} (\bibinfo {year} {2016})}\BibitemShut {NoStop}%
\bibitem [{\citenamefont {Farías}\ \emph {et~al.}(2013)\citenamefont
  {Farías}, \citenamefont {Quinteiro},\ and\ \citenamefont
  {Tamborenea}}]{farias_photoexcitation_2013}%
  \BibitemOpen
  \bibfield  {author} {\bibinfo {author} {\bibfnamefont {M.~B.}\ \bibnamefont
  {Farías}}, \bibinfo {author} {\bibfnamefont {G.~F.}\ \bibnamefont
  {Quinteiro}}, \ and\ \bibinfo {author} {\bibfnamefont {P.~I.}\ \bibnamefont
  {Tamborenea}},\ }\href {\doibase 10.1140/epjb/e2013-40621-2} {\bibfield
  {journal} {\bibinfo  {journal} {The European Physical Journal B}\ }\textbf
  {\bibinfo {volume} {86}},\ \bibinfo {pages} {432} (\bibinfo {year}
  {2013})}\BibitemShut {NoStop}%
\bibitem [{\citenamefont {Ishikawa}()}]{ishikawa}%
  \BibitemOpen
  \bibfield  {author} {\bibinfo {author} {\bibfnamefont {K.~L.}\ \bibnamefont
  {Ishikawa}},\ }\href {\doibase 10.1103/PhysRevB.82.201402} {\ \textbf
  {\bibinfo {volume} {82}},\ \bibinfo {pages} {201402}}\BibitemShut {NoStop}%
\bibitem [{\citenamefont {Katsnelson}()}]{grapheneBook}%
  \BibitemOpen
  \bibfield  {author} {\bibinfo {author} {\bibfnamefont {M.~I.}\ \bibnamefont
  {Katsnelson}},\ }\href {\doibase 10.1017/CBO9781139031080} {\emph {\bibinfo
  {title} {Graphene: Carbon in Two Dimensions}}}\ (\bibinfo  {publisher}
  {Cambridge University Press})\BibitemShut {NoStop}%
\bibitem [{\citenamefont {Barnett}\ and\ \citenamefont
  {Radmore}()}]{methodsQO}%
  \BibitemOpen
  \bibfield  {author} {\bibinfo {author} {\bibfnamefont {S.}~\bibnamefont
  {Barnett}}\ and\ \bibinfo {author} {\bibfnamefont {P.}~\bibnamefont
  {Radmore}},\ }\href {\doibase 10.1093/acprof:oso/9780198563617.001.0001}
  {\emph {\bibinfo {title} {Methods in Theoretical Quantum Optics}}}\ (\bibinfo
   {publisher} {Oxford University Press})\BibitemShut {NoStop}%
\bibitem [{\citenamefont {Haug}\ and\ \citenamefont {Koch}(2009)}]{koch}%
  \BibitemOpen
  \bibfield  {author} {\bibinfo {author} {\bibfnamefont {H.}~\bibnamefont
  {Haug}}\ and\ \bibinfo {author} {\bibfnamefont {S.~W.}\ \bibnamefont
  {Koch}},\ }\href@noop {} {\emph {\bibinfo {title} {Quantum theory of the
  optical and electronic properties of semiconductors}}}\ (\bibinfo
  {publisher} {World Scientific Publishing Company},\ \bibinfo {year}
  {2009})\BibitemShut {NoStop}%
\bibitem [{\citenamefont {Katsnelson}(2007)}]{katsnelson}%
  \BibitemOpen
  \bibfield  {author} {\bibinfo {author} {\bibfnamefont {M.~I.}\ \bibnamefont
  {Katsnelson}},\ }\href@noop {} {\bibfield  {journal} {\bibinfo  {journal}
  {Materials today}\ }\textbf {\bibinfo {volume} {10}},\ \bibinfo {pages} {20}
  (\bibinfo {year} {2007})}\BibitemShut {NoStop}%
\bibitem [{\citenamefont {Kuklinski}\ and\ \citenamefont
  {Mukamel}(1991)}]{refN}%
  \BibitemOpen
  \bibfield  {author} {\bibinfo {author} {\bibfnamefont {J.}~\bibnamefont
  {Kuklinski}}\ and\ \bibinfo {author} {\bibfnamefont {S.}~\bibnamefont
  {Mukamel}},\ }\href@noop {} {\bibfield  {journal} {\bibinfo  {journal}
  {Physical Review B}\ }\textbf {\bibinfo {volume} {44}},\ \bibinfo {pages}
  {11253} (\bibinfo {year} {1991})}\BibitemShut {NoStop}%
\bibitem [{\citenamefont {Kelardeh}\ \emph {et~al.}(2015)\citenamefont
  {Kelardeh}, \citenamefont {Apalkov},\ and\ \citenamefont {Stockman}}]{ref4}%
  \BibitemOpen
  \bibfield  {author} {\bibinfo {author} {\bibfnamefont {H.~K.}\ \bibnamefont
  {Kelardeh}}, \bibinfo {author} {\bibfnamefont {V.}~\bibnamefont {Apalkov}}, \
  and\ \bibinfo {author} {\bibfnamefont {M.~I.}\ \bibnamefont {Stockman}},\
  }\href@noop {} {\bibfield  {journal} {\bibinfo  {journal} {Physical Review
  B}\ }\textbf {\bibinfo {volume} {91}},\ \bibinfo {pages} {045439} (\bibinfo
  {year} {2015})}\BibitemShut {NoStop}%
\bibitem [{\citenamefont {Mandel}\ and\ \citenamefont {Wolf}()}]{mandelWolf}%
  \BibitemOpen
  \bibfield  {author} {\bibinfo {author} {\bibfnamefont {L.}~\bibnamefont
  {Mandel}}\ and\ \bibinfo {author} {\bibfnamefont {E.}~\bibnamefont {Wolf}},\
  }\href {\doibase 10.1017/CBO9781139644105} {\emph {\bibinfo {title} {Optical
  Coherence and Quantum Optics}}},\ \bibinfo {edition} {1st}\ ed.\ (\bibinfo
  {publisher} {Cambridge University Press})\BibitemShut {NoStop}%
\bibitem [{\citenamefont {Stroucken}\ \emph {et~al.}(2011)\citenamefont
  {Stroucken}, \citenamefont {Gr{\"o}nqvist},\ and\ \citenamefont
  {Koch}}]{koch2}%
  \BibitemOpen
  \bibfield  {author} {\bibinfo {author} {\bibfnamefont {T.}~\bibnamefont
  {Stroucken}}, \bibinfo {author} {\bibfnamefont {J.}~\bibnamefont
  {Gr{\"o}nqvist}}, \ and\ \bibinfo {author} {\bibfnamefont {S.}~\bibnamefont
  {Koch}},\ }\href@noop {} {\bibfield  {journal} {\bibinfo  {journal} {Physical
  Review B}\ }\textbf {\bibinfo {volume} {84}},\ \bibinfo {pages} {205445}
  (\bibinfo {year} {2011})}\BibitemShut {NoStop}%
\bibitem [{{\relax DLMF}()}]{nist}%
  \BibitemOpen
  {\relax DLMF},\ \href {http://dlmf.nist.gov/} {\enquote {\bibinfo {title}
  {{\it NIST Digital Library of Mathematical Functions}},}\ }\bibinfo
  {howpublished} {http://dlmf.nist.gov/, Release 1.1.5 of 2022-03-15},\
  \bibinfo {note} {f.~W.~J. Olver, A.~B. {Olde Daalhuis}, D.~W. Lozier, B.~I.
  Schneider, R.~F. Boisvert, C.~W. Clark, B.~R. Miller, B.~V. Saunders, H.~S.
  Cohl, and M.~A. McClain, eds.}\BibitemShut {Stop}%
\end{thebibliography}%
\end{document}